# Anomalous switching in Nb/Ru/Sr$_2$RuO$_4$ topological junctions by chiral domain wall motion

M. S. Anwar[1,*], Taketomo Nakamura[1,2], S. Yonezawa[1], M. Yakabe[3], R. Ishiguro[3], H. Takayanagi[3,4] and Y. Maeno[1]

[1]Department of Physics, Kyoto University, Kyoto 606-8502, Japan

[2]Institute for Solid State Physics, the University of Tokyo, Kashiwa 277-8581, Japan

[3]Department of Applied Physics, Faculty of Science, Tokyo University of Science, Tokyo 162-8601, Japan

[4]International Center for Materials Nanoarchitectonics (MANA), National Institute for Materials Science (NIMS), Tsukuba 305-0044, Japan

**A spontaneous symmetry breaking in a system often results in domain wall formation. The motion of such domain walls is utilized to realize novel devices like racetrack-memories, in which moving ferromagnetic domain walls store and carry information. Superconductors breaking time reversal symmetry can also form domains with degenerate chirality of their superconducting order parameter. Sr$_2$RuO$_4$ is the leading candidate of a chiral *p*-wave superconductor, expected to be accompanied by chiral domain structure. Here, we present that Nb/Ru/Sr$_2$RuO$_4$ topological superconducting-junctions, with which the phase winding of order parameter can be effectively probed by making use of real-space topology, exhibit unusual switching between higher and lower critical current states. This switching is well explained by chiral-domain-wall dynamics. The switching can be partly controlled by external parameters such as**

**temperature, magnetic field and current. These results open up a possibility to utilize the superconducting chiral domain wall motion for future novel superconducting devices.**

Since the discovery of superconductivity in $Sr_2RuO_4$ (SRO)[1], various experiments[2-7] reveal that the pairing state of SRO is of chiral *p*-wave spin-triplet with broken time reversal symmetry[8,9], analogous to the A-phase of superfluid $^3$He[10,11]. However, the issue of chiral *p*-wave nature of SRO still remains controversial since some of the predicted behavior such as chiral edge current has not been observed[9]. Thus establishment of novel behavior specific to chiral *p*-wave superconductivity is much desirable. Recently, SRO is considered as one of the most promising materials for exploring topological superconducting phenomena originating from its orbital phase winding[9]. Because of nontrivial topological aspect of its superconducting order parameter, gapless chiral edge states consisting of Majorana quasiparticles (whose antiparticles are their own particles) are believed to emerge at its boundaries[12-15].

Chiral *p*-wave superconductivity exhibits two-fold degeneracy corresponding to clockwise or counterclockwise winding of the superconducting phase. This degeneracy sets up two kinds of chiral domains separated by a chiral domain wall (chiral-DW)[16]. To date, there is no direct observation of the chiral-DW[17]. However, the existence of the chiral-DW has been strongly suggested by transport studies of SRO-based junctions[2,18]. Further accumulation of evidence of the chiral-DW and investigations of possible influences of chiral-DW dynamics on transport properties are important because the chiral-DW can be utilized for novel superconducting devices as in the case of the ferromagnetic-DW for racetrack memory devices[19].

A "topological junction" consists of a superconductor surrounded by another in such a way that the difference in phase winding dictates the junction behavior[20,21]. The characteristics of a topological junction with a chiral *p*-wave superconductor should be very sensitive to the chiral domain configuration. The SRO-Ru eutectic system[22] provides naturally existing topological junctions, once *s*-wave superconductivity is induced into Ru inclusions surrounded by SRO. Indeed, junctions fabricated using Pb as an *s*-wave superconductor deposited over many Ru-inclusions[20,21] exhibit peculiar temperature dependence of critical current $I_c$ attributable to topological phase competition between the *s*-wave and *p*-wave superconductivity. Since previous devices containing many Ru junctions probe only averaged effects, it is much desirable to fabricate a device with a single junction to investigate the order parameter structure more sensitively, including the effect of chiral domains.

**Results**

We fabricate SRO-Ru based micron-sized junctions utilizing only one Ru inclusion shown in Figs. 1a-d. Figure 1e presents the junction resistance versus temperature. Junction A (Junction B) exhibits the first transition at 9.5 K (9 K) corresponding to the superconducting transition temperature $T_c$ of Nb. The final transition starting at around 2.8 K (for both junctions) leads to zero junction resistance at $T_{c,A} = 1.68$ K ($T_{c,B} = 1.86$ K) (inset of Fig. 1e). These temperatures are significantly higher than $T_{c\_bulk} = 1.42$ K of SRO in the eutectic crystal used in this study because of enhanced superconductivity at the interface between Ru and SRO, the so-called 3-K phase[22]. A clear supercurrent branch with zero voltage is obvious in an *I-V* curve at 0.37 K (inset of Fig. 2a). These facts, as well as Fraunhofer pattern (see the supplementary information), indicate that our junctions exhibit a typical Josephson coupling.

Figure 2a presents $I_c$ versus temperature data, accumulated from 0.34 K to 2.5 K with various cooling cycles (represented with different colors). Interestingly, we find a sharp jump in $I_c$ after "each" cooling cycle. Such jumps are prominent at $T < T_{c\_bulk}$; at $T > T_{c\_bulk}$ $I_c$ is rather stable. The changes in $I_c$ indicate the switching between two states of the junction with the cooling cycles.

To study the $I_c$ variations further, we obtained *I-V* curves at various temperatures after zero-field cooling (Fig. 3a). An ordinary *I-V* curve is observed at 1.5 K. However, *I-V* curves at 1.4 K and 0.5 K exhibit oscillations between zero and non-zero voltages corresponding to switching between higher-$I_c$ and lower-$I_c$ states. At these three temperatures the voltage versus time $V(t)$ is also recorded at constant excitation current $I_{exc}$ just below $I_c$ (Fig. 3b). At 1.5 K, $V(t)$ exhibits constant zero voltage. However, at 1.4 K, $V(t)$ shows sharp switching between zero and non-zero voltages ≈120 nV. This switching resembles telegraphic noise (TN). Nearly equal probabilities in the non-zero and zero voltage states indicate that the lower-$I_c$ state is as stable as the higher-$I_c$ state. The $V(t)$ data at 0.5 K demonstrate rather sharp and short switching at the amplitude of ~200 nV. Thus, the junction tends to stay in the higher-$I_c$ state. These observations, as well as sudden disappearance of TN signal at $T_{c\_bulk}$ in data (see the supplementary information) taken under a temperature up-sweep, reveal that the switching behavior is strictly correlated with the bulk superconductivity in SRO; the junction is quite stable at $T > T_{c\_bulk}$ and rather unstable at $T < T_{c\_bulk}$. Note that the switching is also observed at different temperatures between 1.4 K and 0.5 K. Although the transition temperature of bulk Ru is 0.49 K, we do not observe any anomaly in $I_c(T)$ at a corresponding temperature (see fig. 2); this observation indicates that Ru is already fully proximitized.

We also preformed experiments to control the switching behavior. Figure 4a shows the influence of $I_{exc}$ at 1.4 K. The $V(t)$ data at $I_{exc} = 30$ µA, about half of $I_c = 62$ µA, show zero resistance. At $I_{exc} = 53$ µA, $V(t)$ exhibits voltage switching of the order of 120 nV. Note that switching between the high voltage state and an intermediate state is sometimes observed. Overall, a longer time in the zero-voltage state indicates that the junction is more stable in the higher-$I_c$ state. Closer to $I_c$ ($I_{exc} = 59$ µA), the junction spends nearly equal time in both states. For $I_{exc} = 65$ µA $> I_c$, $V(t)$ exhibits constant non-zero voltage. Thus, the switching is only observed in the $I_{exc}$ range where the voltage oscillations are present in the corresponding $I$-$V$ curves (Fig. 3a). Close to $I_c$, where the junctions are rather unstable, we found that a small temperature variation can trigger the switching: in the $V(t)$ curve at 1.4 K with intentional temperature variations of ~1.5 mK, the switching occurs in-phase to the temperature variations (Fig. 4b). Note that the temperature variations during $V(t)$ measurements of the curves in Figs. 3&4a were smaller than 50 µK; this fact evidences that the temperature variations can stimulate the switching but is not the origin. We also found that switching is enhanced by small externally applied magnetic fields. Figure 4c shows the $I$-$V$ curve at 0.5 K with the field of 0.10 Oe along the *ab*-plane exhibiting fine voltage variations, with the corresponding $V(t)$ data also showing fast switching (inset of Fig 4c).

We also demonstrate that switching behavior can be altered by cooling cycles. Figure 3 and Fig 4d show data at different cooling cycles at 0.5 K. The switching, which is obvious in the former case, is not observed in the latter case. The switching is not observed for $I_{exc} < I_c$ either (upper-left inset of Fig. 4d). It is interesting that the hysteresis loop in *I-V* curves for the latter case is reversed, in the sense that zero voltage state is realized with higher current for down sweep. This hysteretic behavior is also anomalous because the difference between the higher and lower $I_c$ is not constant; sometimes the hysteresis in $I_c$ disappears (bottom-right inset of Fig. 4d). These facts reveal that the system in this cooling cycle is rather stable but anomalous hysteresis suggests that some instability is present at $I_{exc} > I_c$. Indeed, we observed a small switching only at $I_{exc} = 144$ µA ($I_{exc} > I_c$), but surprisingly not for $I_{exc} = 143$ or 145 µA (see the supplementary information).

**Discussion**

Prior to discussion, we summarize the behavior of the Nb/Ru/SRO junctions. With decreasing temperature below $T_{c\_Nb} = 9.5$ K, the proximity effect of the *s*-wave superconductivity develops in Ru. Below 3 K, the interfacial 3-K superconductivity in SRO sets in and the junctions start to show finite $I_c$ below ~1.8 K, forming SNS' junctions. Although the junction behavior is conventional and highly reproducible down to $T_{c\_bulk}$, a number of anomalous behaviors emerge at temperatures precisely below $T_{c\_bulk}$. First is the anomalous hysteresis in the *I-V* curves, often accompanied by asymmetry with respect to the direction of current. A similar hysteresis has been reported by Kambara *et al.*, in SRO-Ru micro-bridge[18]. Second is the presence of mainly two branches of $I_c$, between which junctions switch back and forth. Third is the TN, which corresponds to the telegraphic switching between the multiple branches of $I_c$. The junctions at temperatures just below $T_{c\_bulk}$ show rather active TN with mainly two different states. At low temperature the junctions are more stable in the higher-$I_c$ state. Whenever the TN is active $I_c$ drops down by ≈50% to $I_c$ in the most stable state. The junctions can be driven into unstable state with active TN either by different cooling cycle or by tiny external magnetic fields. Comparing Junctions A and B, Junction A with smaller junction area is relatively stable. These behaviors cannot be explained by the motion of ordinary vortex (see the supplementary information). Below, we examine the possible origins of the unusual behavior in terms of self-induced vortex dynamics specific to chiral superconductor, and in terms of chiral-DW dynamics.

Self-induced vortex: For a Ru-inclusion below its $T_c$ (0.49 K) surrounded by SRO, it is theoretically predicted that a self-induced vortex appears due to a competition between $s$-wave superconductivity in Ru metal and the chiral $p$-wave superconductivity in SRO. Such a vortex can switch between two states: one at the Ru/SRO interface and the other at the center of the Ru-inclusion. The switching should occur at lower temperatures and more likely for smaller Ru-inclusions[23]. In our junctions, similar self-induced vortex is anticipated even above 0.5 K because of the proximity-induced superconductivity in Ru via Nb electrode. However, our junctions become stable at lower temperatures and also for smaller Ru-inclusion. Thus, the observed behavior is unlikely caused by the self-induced vortex dynamics.

Chiral-DW dynamics: Chiral domain structure of superconducting bulk SRO around the Ru-inclusion is expected to play a crucial role in determining $I_c$. It is considered to appear only below $T_{\text{c\_bulk}}$[20,21], in accord with the emergence of unusual behavior observed only below $T_{\text{c\_bulk}}$. Expectedly, it becomes more stable at low temperatures with increasing condensation energy of SRO. To illustrate the effect of chiral-DW dynamics, let us introduce a simplest model. We consider a single Ru-inclusion having a smooth circular-shape in pure SRO and two chiral-DWs separating the imbedded in SRO into two domains with opposite chirality ($\eta_+ = e^{i\theta}$ and $\eta_- = e^{-i\theta}$, Fig. 5a). The chiral-DWs intersect with the Ru/SRO interface fixed at $\theta(\text{F}) = 0°$ with $\varphi_+(\text{F}) = \Delta\varphi$ ($\theta$ is the direction normal to the interface, $\varphi_\pm$ is the phase difference of the p-wave SRO in the domain $\eta_\pm$ with respect to the s-wave Ru) and at $\theta(\text{M}) = \theta_{\text{DW}}$ which is assumed movable. The free energy of a chiral-DW depends on the orientation and the phase difference $\alpha$ across the chiral-DW, which would be determined by minimizing the junction energy for a given "M" and $I_{\text{exc}}$ with the distribution of the pinning potential for chiral-DWs, etc. in an intricate way[16]. Here, the effect of the additional energy associated with the induced magnetic flux is not included. This is because the phase winding mismatch between s-wave and p-wave is resolved by presence of a chiral domain wall and as a consequence the flux energy is expected to be much reduced. Thus, we introduce a conjecture that $I_c$ is determined by the following current-phase relation for a given $\theta_{\text{DW}}$ which is varied from 0 to π,

$$I_c = \max\left[\frac{I_{c0}}{2\pi}\left\{\int_0^{\theta_{\text{DW}}} d\theta \sin\varphi_+(\theta; \theta_{\text{DW}}, \Delta\varphi) + \int_{\theta_{\text{DW}}}^{2\pi} d\theta \sin\varphi_-(\theta; \theta_{\text{DW}}, \Delta\varphi)\right\}\right]. \quad (1)$$

By symmetry and for single-valuedness of the order parameter we consider $\alpha(\text{M}) = -\alpha(\text{F}) = \pi - \theta_{\text{DW}}$. This in fact is the condition to maximize $I_c$ for a given value of $\theta_{\text{DW}}$. In this scenario, maximum $I_c$ is realized at $\theta_{\text{DW}} = \pi$ and $\alpha = \Delta\varphi = 0$ (Fig. 5b). This can be understood as the absence of current cancelation over the Ru circumference in the presence of multiple chiral-DWs (see the supplementary information). A rotation of domain wall M by ±10º around $\theta_{DW} = \pi$ affects $I_c$ very little, but the rotation around $\theta_{DW} = \pi/2$ significantly changes $I_c$ (Fig. 5c; *I-V* curves are calculated using the relation for an overdamped junction). This character captures features of our experimental observations (Fig. 5d): stable and maximum $I_c$ is observed in one cooling cycle, and unstable and lower $I_c$ is realized in a different cooling cycle. In reality, it is reasonable that $\theta_{DW} = \pi$ gives the most stable state because the actual Ru inclusion is elongated with the maximum curvature (maximum disorder) at its corners providing maximum pinning for chiral-DWs. Note that the observed $I_c$ is always ~50% lower when voltage oscillations emerge in the *I-V* curves. In our model such lower and unstable $I_c$ occurs for the chiral-DW motion around $\theta_{DW} = \pi/2$, corresponding to the flat part of the actual Ru/SRO interface. These calculations confirm that the aspects of the observed anomalous behavior of our junctions are well explained by the chiral-DW motion.

We studied $Sr_2RuO_4$-based micron-sized junctions, Nb/Ru/SRO, using one Ru inclusion, and found unusual temperature dependence of $I_c$, anomalous hysteresis with current, and switching in $I_c$. It is difficult to explain the overall behavior by vortex dynamics (ordinary and self-induced). Instead, a simple model based on the chiral-DW motion captures the main features of the observed junction behavior. Our results provide further evidence for chiral *p*-wave order parameter in SRO and reveal the crucial effects of chiral-DW motion on Josephson coupling. The switching raised by chiral-DW motion can be controlled by various external parameters and provides a ground for novel superconducting devices, analogous to memory devices based on ferromagnetic-DW motion. Our work also demonstrates the scientific importance of the concept of the topological junctions to expose the phase winding of superconducting order parameter by making use of the real-space topology.

**Methods**

We fabricated Nb/Ru/SRO micron-sized Josephson junctions using a polished (the basal *ab*-plane) rectangular pieces (3×3×0.5 mm$^3$) of SRO-Ru eutectic crystals grown by a floating-zone method[24]. Contact resistance between the *ab*-surface of SRO and Nb is rather large but Ru metal works as an adhesive layer to provide a good contact. Although, a technique of using Nb/Cu bilayer has recently been developed to establish a good contact to the surface parallel to the *c*-axis to enhance the $J_c$ of the junction[25], here we need to deal with the *ab*-plane surface contact. After polishing its *ab*-surface, SiO$_x$ layer of thickness of ~300 nm was deposited using RF sputtering technique with a backing pressure of ~10$^{-7}$ mbar. Then a photoresist (TSMR-8800) was coated, exposed with laser lithography over only one Ru inclusion. The exposed resist was removed with TMAH2.83% developer for 120 sec followed by rinsing in DI-water for 30 sec and dried with N$_2$ gas. A part of the SiO$_x$ film covering single Ru inclusion was etched with CHF$_3$ gas, which opened the windows over a single Ru inclusion (Fig. 1a). In this process, a fluoride thin film may be generated on the surface of the sample. We performed an O$_2$ plasma cleaning step to etch away a fluoride film. The resist was removed using N-Methyl-2-pyrolidone (NMP) and cleaned with acetone and isopropanol. In the next step to deposit the Nb electrodes, we used a lift-off technique using bilayer photoresist (LOR-10A and TSMR-8800) and laser lithography.

A Nb film of the thickness of ~1 μm was sputtered with a base pressure of ~$10^{-7}$ mbar. Finally, the lift-off was accomplished with NMP. Note that Nb is not only in contact with Ru but also with SRO along *ab*-plane. It is well known that *ab*-plane is less conductive because of atomic reconstruction at the surface and does not allow supercurrent to flow directly from Nb. Instead, supercurrent from Nb passes only through the Ru metal with proximity induced superconductivity from Nb. Figure 1b shows an overall scanning electron microscope (SEM) picture of the device with two junctions. We measured the *I-V* curves using four-point technique with two contacts at Nb over the *ab*-plane and the other two contacts on the side directly connected via silver paste with SRO crystal as shown in the schematic of the side view in Fig. 1c. The measurements are performed with a $^3$He cryostat down to 300 mK. The cryostat was magnetically shielded with high-permeability material (Hamamatsu Photonics, mu-metal). Inside the shield, we placed a superconducting magnet to apply the magnetic fields.

**Acknowledgement:** We are grateful to S. Kashiwaya and D. Manske for fruitful discussions. We are thankful to Y. Yamaoka and Y. Nakamura for valuable discussions and help in the measurements. This work was supported by the "Topological Quantum Phenomena" (No. 22103002 and 251037021) KAKENHI on Innovative Areas from Ministry of Education, Culture, Sports, Science and Technology (MEXT) of Japan. This study was also supported by NIMS Nanofabrication Platform in "Nanotechnology Platform Project" sponsored by MEXT, and we would like to thank E. Watanabe, D. Tsuya and H. Oosato for technical supports. This work was also supported by the Japan Society for Promotion of Science (JSPS) KAKENHI S (No. 20221007).

**Authors Contributions:** M. S. A. did the measurements and model calculations. T. N. prepared the SRO crystals, design the devices, and measured $R(T)$. S. Y. was also involved the measurements. M. Y., R. I. and H. T. fabricated the devices. Y. M. designed and supervised the work. M.S.A., S.Y. and Y.M. analyzed the data and wrote the article. All the authors were involved in the discussion and reviewed the manuscript.

**Additional information**

Competing financial interests: The authors declare no competing financial interests.

**Figure Captions**

**Figure 1| Topological junction devices and the resistance versus temperature. a,** Optical microscope image of the Nb contact area with SRO over a single Ru inclusion of cross sectional area 1×5 μm$^2$ for junction A (shown in the inset) and 1×14 μm$^2$ for Junction B. The elongated bright part in the blue box (where SiO$_x$ layer is etched out) is the Ru inclusion and blue indicates SRO. **b,** Scanning Electron Microscope (SEM) image of the Nb/Ru/SRO device with two junctions A and B. White dotted lines show the boundary of Nb electrodes and black rectangular areas indicate the positions of the junctions. Note that the contrast between Ru and SRO is clear under optical microscope but not under SEM. That is the reason we used optical microscope for the zoomed-in image. **c,** 3D schematic view of a junction. **d,** Schematic cross section of the junction, fabricated by depositing ~1 μm thick Nb electrode after depositing ~300 nm thick SiO$_x$ layer with an open window over a single Ru inclusion. **e,** Resistance versus temperature for both Junctions A and B. The normal junction resistance $R_N$ is 128 mΩ for the junction A (blue curve) and 11.5 mΩ for the junction B (red curve). The different $R_N$ values are attributed to different interface transparency as well as cross sectional area. The drop in the resistance at around 9 K corresponds to superconducting transition of Nb ($T_c \approx 9.5$ K) and the drop to zero resistance starts at 2.8 K because of proximity effect with 3-K phase. There are additional drops for the junction A, reflecting gradual development of proximity into Ru metal. The inset is an enlargement showing that the zero junction resistance persists to temperatures substantially above $T_{c\_bulk} = 1.42$ K of SRO. The bulk resistance of SRO is negligible compared with the junction resistance.

**Figure 2 | Temperature dependent critical current of the topological junctions.** Critical current $I_c$ as a function of temperature $T$ at zero field for Junctions B. During the measurements the temperature was raised above 3 K several times due to the liquid $^3$He hold time of our refrigerator. To continue our measurements we cooled it down again. In this way the $I_c(T)$ data were accumulated with various cooling cycles: different colors represent cooling cycles. There are sharp $I_c$ jumps that start with every subsequent cooling cycle, indicating two different $I_c$ braches. The dotted red curves illustrate two different $I_c$ branches. Inset shows a current-voltage (*I-V*) curve at 0.37 K for Junction B. It shows a clear zero voltage supercurrent up to ~450 μA in the most stable state.

**Figure 3 | Switching behavior in topological junctions. a,** *I-V* curves at various temperatures for Junction B at zero field after zero-field cooling. The *I-V* curve at 1.5 K exhibits ordinary behavior. However, at 1.4 K, and 0.5 K the *I-V* curves show switching between zero and non-zero voltages near the transition region. The inset presents the enlarged *I-V* curves at 1.5 K and 1.4 K, clearly illustrating that the switching occurs only at 1.4 K. This fact indicates that the switching is strictly connected with the bulk superconductivity of SRO below 1.42 K. **b,** Voltage versus time $V(t)$ with external excitation current $I_{exc}$ slightly lower than $I_c$ recorded just after measuring the corresponding *I-V* curves given in the panel **a**. At 1.5 K, $V(t)$ (black; voltage is shifted by 500 nV) is constant with the small extrinsic voltage drift of about 10 nV over a period of $10^3$ sec. At 1.4 K, $V(t)$ (red; voltage shifted by 300 nV) shows the sharp switching between the lower-$I_c$ and higher-$I_c$ states. At lower temperature of 0.5 K, $V(t)$ (blue) exhibits similar switching but the junction comes back to the higher-$I_c$ state after a very short time.

**Figure 4 | Influence of different external parameters on switching. a,** $V(t)$ at 1.4 K measured at different $I_{exc}$ values. At $I_{exc}$ =30 µA <$I_c$ ≈62 µA, $V(t)$ (black) represents the zero voltage state. At $I_{exc}$ =53 µA (blue; shift of 70 nV for clarity) switching between lower and higher-$I_c$ states occurs. At $I_{exc}$ =59 µA close to $I_c$ (red; shift of 200 nV), the switching occurs more frequently and the junction tends to spend more time in the lower-$I_c$ state. At $I_{exc}$ =65 µA>$I_c$ (green: shift of 80 nV) the voltage is constant at ≈300 nV. **b,** $I_c$ switching triggered by temperature variations (≈1.5 mK). $V(t)$ at 1.4 K shows in-phase voltage oscillations with temperature variations (blue arrows). Note that transition to the lower voltage state is sometimes absent (green arrows). It suggests that temperature variations can control the voltage oscillations but is not the origin. **c,** External magnetic-field effect on switching. The $I$-$V$ curve was obtained under a field of 0.10 Oe at 0.5 K after zero-field cooling. Although $I_c$ is suppressed, switching still exists with smaller period of time. $V(t)$ in the inset also indicates fast switching. **d,** Cooling cycle effect on switching at 0.5 K. $I$-$V$ curves were obtained in zero field after zero-field cooling. It shows anomalous hysteresis without switching, in contrast to the case shown in Fig. 3a recorded in different cooling cycle. The switching in $V(t)$ is also absent (upper-left inset), when only the hysteresis is present. The hysteresis can be sometimes absent (lower-right inset).

**Figure 5| $I_c$ switching originating from chiral domain wall motion. a,** Schematic of Sr$_2$RuO$_4$ sample with one Ru inclusion. The figure represents the NS' part of an SNS' (Nb/Ru/Sr$_2$RuO$_4$) junction, where N (Ru-metal) is proximitized by conventional $s$-wave superconductor S (Nb). Two chiral domain walls intersecting the interface between Ru and SRO are also shown. One domain wall "F" is considered to be fixed at angle $\theta = 0$ with the

phase difference between the *s*-wave and the *p*-wave superconductors $\varphi_+(\text{F}) = \Delta\varphi$. The second domain wall "M" is considered to be movable and located at $\theta = \theta_{\text{DW}}$ with $\varphi_+(\text{M}) = \theta_{\text{DW}} + \Delta\varphi$. For the maximum critical current for a given value of $\theta_{\text{DW}}$, phase differences across the domain walls are $\alpha(\text{M}) = -\alpha(\text{F}) = \pi - \theta_{\text{DW}}$. The color wheels represent the evolution of the superconducting phase $\varphi$ with the direction $\theta$. **b,** Calculated $I_c$, $\alpha$, and $\Delta\varphi$ as functions of $\theta_{\text{DW}}$. Maximum $I_c$ is realized at $\theta_{\text{DW}} = \pi$ with $\alpha = \Delta\varphi = 0$. **c,** Calculated *I-V* curves for domain wall motion over $\pm 10°$ around $\theta_{\text{DW}} = \pi$ and $\theta_{\text{DW}} = \pi/2$. **d**, Experimentally obtained *I-V* curves at 0.5 K in two different cooling cycles. Red curves are representing many traces of current sweep without changing the temperature.

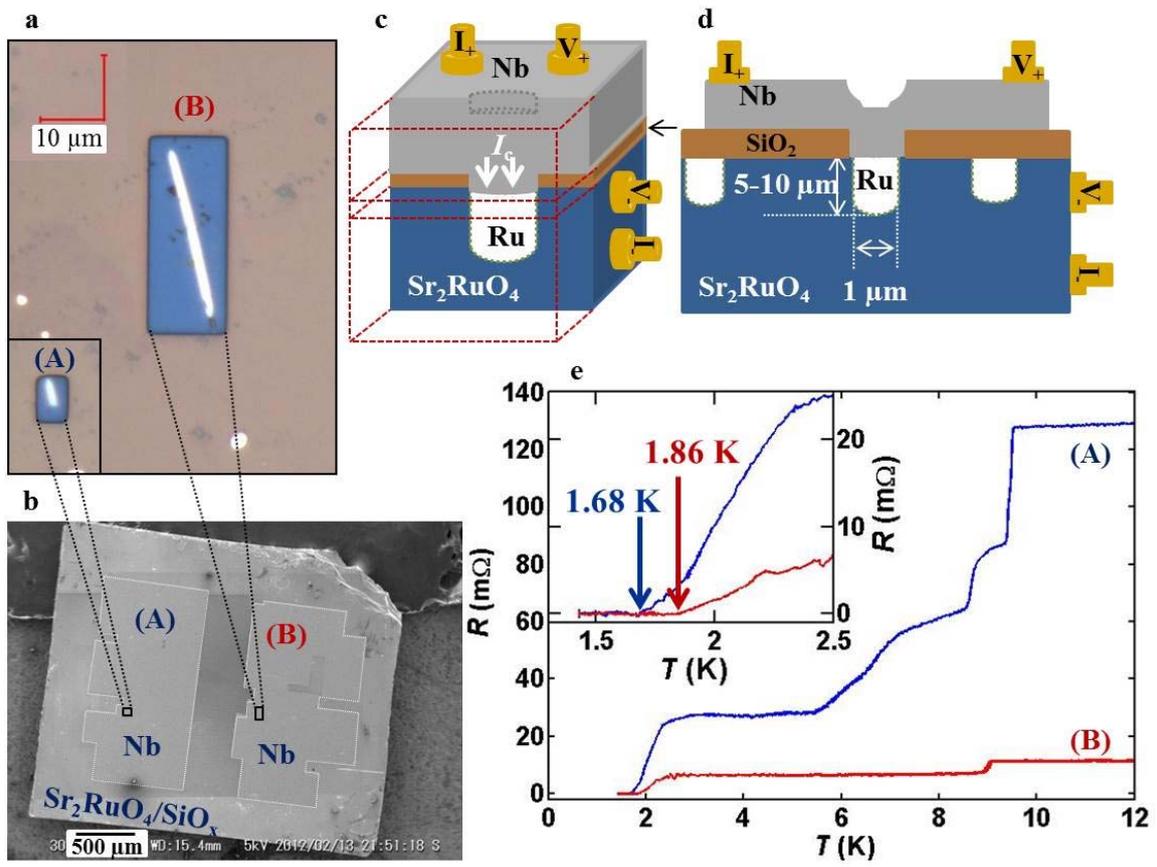

Figure 1

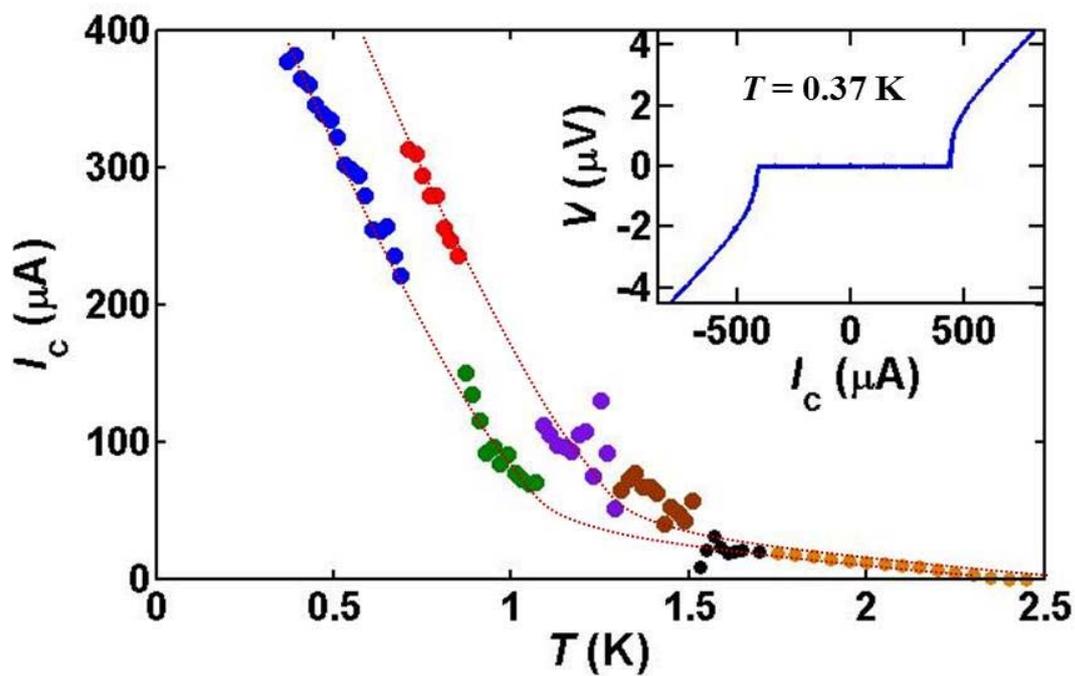

Figure 2

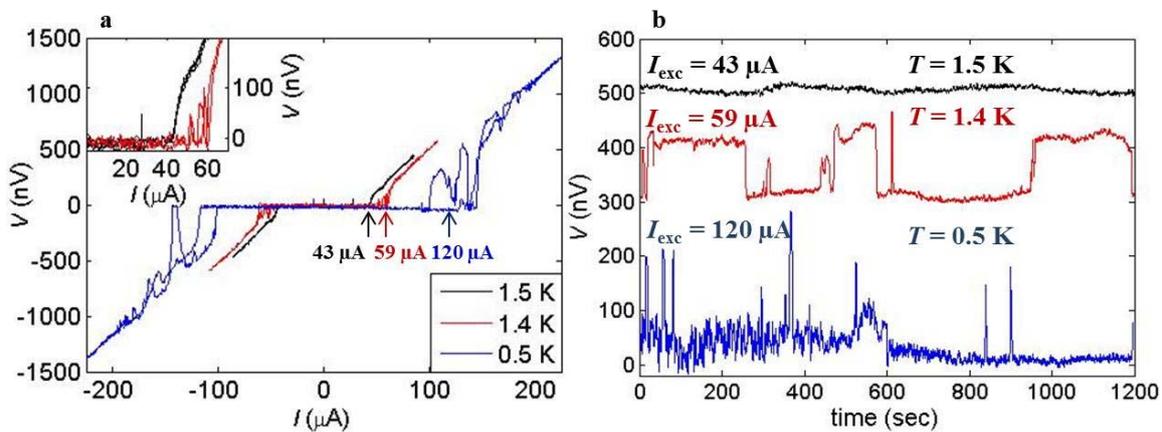

Figure 3

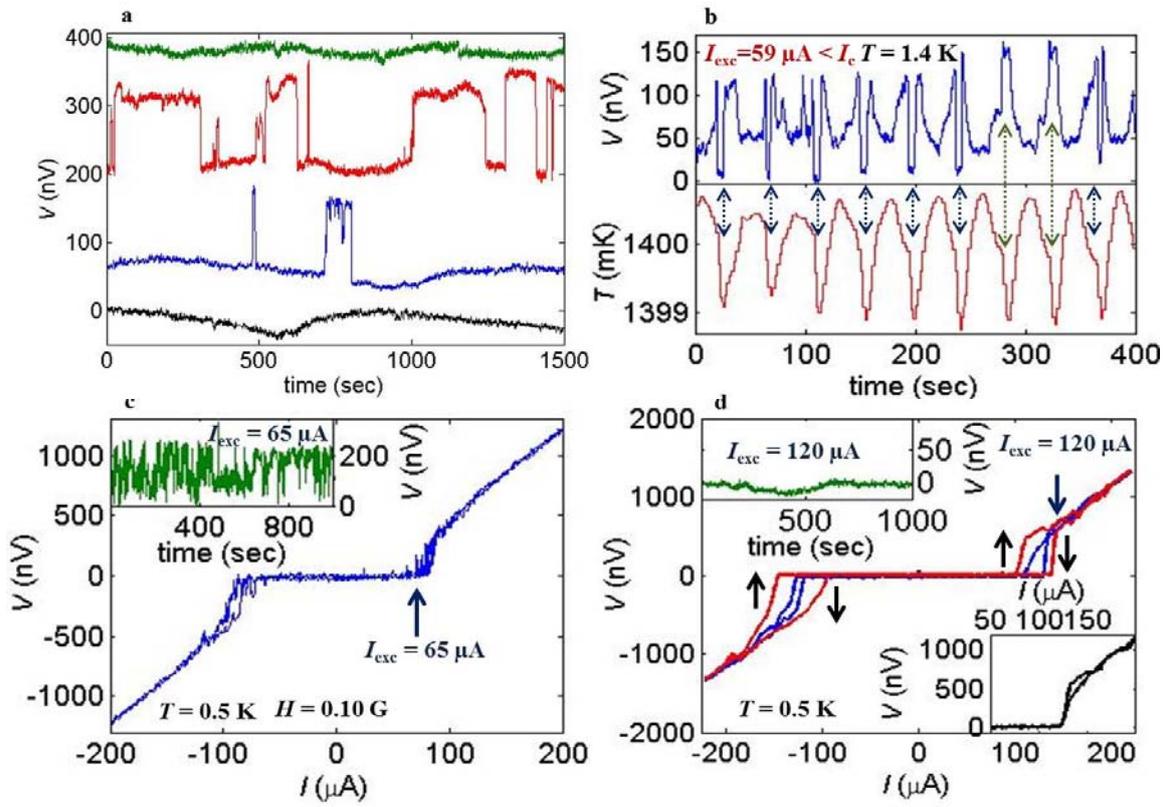

Figure 4

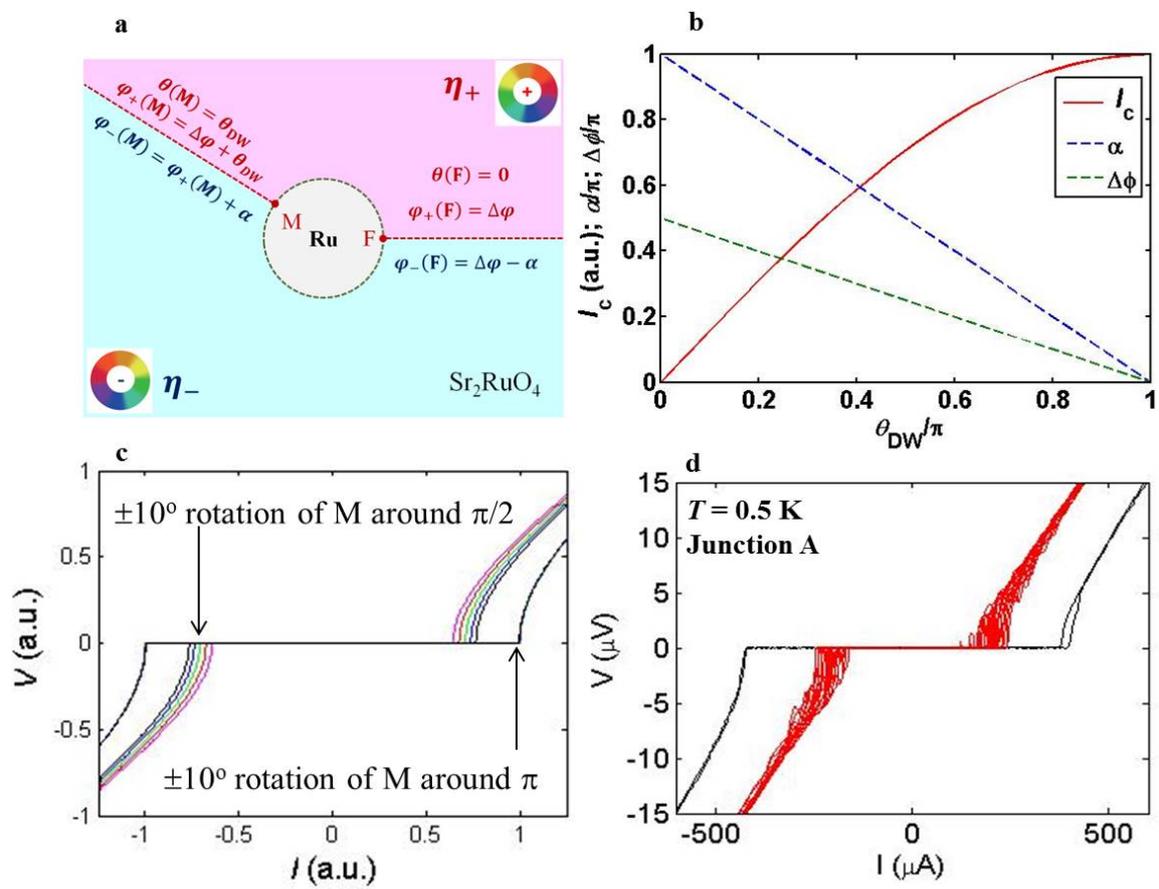

Figure 5

# Supplementary information for

# Anomalous awitching in Nb/Ru/Sr$_2$RuO$_4$ topological junctions by chiral domain wall motion

M. S. Anwar[1], Taketomo Nakamura[1,2], S. Yonezawa[1], M. Yakabe[3],

R. Ishiguro[3], H. Takayanagi[3,4] and Y. Maeno[1]

[1]Department of Physics, Kyoto University, Kyoto 606-8502, Japan

[2]Institute for Solid State Physics, the University of Tokyo, Kashiwa277-8581, Japan

[3] Department of Applied Physics, Faculty of Science, Tokyo University of Science, Tokyo 162-8601, Japan

[4]International Center for Materials Nanoarchitectonics (MANA), National Institute for Materials Science (NIMS), Tsukuba 305-0044, Japan

This supplementary information provides additional data and discussion to elaborate the key results given in the main text. Here, Fraunhofer pattern, asymmetric *I-V* curves, voltage oscillations (telegraphic noise: TN) as a function of temperature, TN at 0.5 K for $I_{exc} > I_c$, and the effect of temperature variations on voltage oscillations are presented. We also discuss further details of the origin of the switching behavior.

Before going into details, we would like to describe characteristic times involved in data acquisition. These characteristic times are important to understand the data, in particular the telegraphic noise. The filtering time and the sampling time of the data acquisition are both much shorter than the transition time, duration, and the frequency of a telegraphic event. We used a Nanovoltmeter Keithley-2182 with the medium speed setting (the filtering time of 50 msec) and set the sampling time of 50 msec. With these settings we observed the transition time of a telegraphic event of the order of 2.5 sec at 1.4 K and 1.8 sec at 0.5 K, which is two orders of magnitude longer than the time of filtering and sampling. The duration of a telegraphic event is of the order of 6.7 sec at 1.4 K and 2.27 sec at 0.5 K. Thus, it is unlikely that we miss a telegraphic event. We cannot be sure that we do not miss telegraphic events above the bulk $T_c$, but it requires that the characteristic times of an event becomes three orders of magnitude shorter above the bulk $T_c$.

SRO is a good metal and has the bulk superconducting transition around 1.42 K. We observe superconducting critical current density $J_c$ above $T_{c\_bulk}$ because of 3-K phase (interfacial superconductivity induced at the interface between Ru and SRO). Very low resistivity of bulk SRO (~ 0.1 µΩcm) gives the normal resistance of the order of µΩ just before the $T_{c\_bulk}$. It reveals that below junction $J_c$ may well contain the normal resistance of the bulk path (~ 0.2 µΩ), which is negligible compared to the junction normal resistance (≈ 7 mΩ for Junction B and 27 mΩ for junction A). What is certain from the data is that below 1.86 K the junction area establishes the SNS' junction with the development of superconductivity at the interface between Ru and SRO that defines the $J_c$ of the junction. That is what revealed by the AC and DC susceptibility measurements, the volume fraction near the $T_{c\_bulk}$ is corresponding to the bulk value [1].

**Fraunhofer pattern**

SNS' junctions made of Nb/Ru/Sr$_2$RuO$_4$ (SRO) exhibit usual behavior of critical current $I_c$ as a function of applied magnetic field termed as Fraunhofer pattern as shown in Fig. S1. The Fraunhofer pattern for Junction B at 0.5 K with applied field along the *ab*-plane shows clear minima at ±130 Oe and some additional shoulders at lower fields. The junction area corresponding to ±130 Oe is ≈ 0.16 µm$^2$. However, the geometry of our topological junction is rather complicated to calculate the junction area which contributes to the $I_c$ modulation. At 0.5 K the proximity penetration depth in the Ru inclusion is of the order of 1 µm. By considering 1 µm long interface between SRO and Ru, which is the main junction, yields the junction depth of the order 160 nm inside the SRO. The additional shoulders suggest mechanism related to dynamical nature of the junction.

**Temperature dependent voltage oscillations**

Our junctions exhibit stable $I_c$ down to the bulk superconducting transition temperature of SRO ($T_{c\_bulk}$) but a number of anomalous features at lower temperatures. Figure S2 shows voltage variations at constant excitation current ($I_{exc}$) as a function of time for Junction A (upper panel); the temperature is gradually increased with time (lower panel). The large voltage variations stop at $T$ = 1.423 K, which corresponds to $T_{c\_bulk}$ of SRO. This fact illustrates that the anomalous behavior emerges from the chiral superconductivity of SRO and is most probably connected with dynamic nature of its chiral domain structure.

**Asymmetric *I-V* curves**

The observed *I-V* curves often become asymmetric with the direction of the current (difference between $I_{c+}$ and $I_{c-}$) for $T < T_{c\_bulk}$ (Fig. S3) but are always symmetric for $T < T_{c\_bulk}$. It is consistent with previous observations Nakamura *et al.*,[2,3]. Note that the sign and magnitude of the observed asymmetry varies with cooling cycles. Asymmetric *I-V* curves are observed for a SQUID consisting on two non-identical junctions. In case of Josephson junction asymmetric *I-V* curves may result from non-identical interfaces. But in both of these cases the asymmetry is persistent with thermal cycles[4]. In such a case, the asymmetry is intrinsic to each junction device and persists in every cooling cycle, in contrast with the present observations. This fact suggests that our junctions have different chiral-DW states during different current sweeps.

**Telegraphic noise at $I_{exc} > I_c$ for relatively stable state**

The TN is mainly observed for $I_{exc}$ lower than but close to $I_c$. Sometimes TN is also observed for $I_{exc} > I_c$. As an example at 0.5 K we present TN at $I_{exc} = 144$ µA $> I_c = 140$ µA in Fig. S4. It is interesting to note that TN is very sensitive to $I_{exc}$: if we increase or decrease $I_{exc}$ just by 1 µA the TN disappears. Voltage is time independent at $I_{exc} = 143$ µA and 145 µA but the voltage value is larger for smaller $I_{exc}$, exhibiting non-ohmic behavior. This observation suggests that the chiral-DW motion may affect the junction resistance in resistive regime beyond $I_c$.

In this situation the reverse hysteresis is also observed. The $I_c R_N$ values for our junctions is in the range of 5-15 µV, which is much smaller than the superconducting gap. It reveals that our junctions are underdamped (very small McCumber parameter) and non-hysteretic *IV*s are expected for conventional junctions. Thus, inverse hysteresis is not coming from the quality of the junction; it might also be related with other mechanisms such as the chiral DW dynamics.

**Influence of temperature variations on the switching**

In the main text we presented the *V*(*t*) data at 1.4 K (Fig. 4b) for which TN are controlled by temperature variations even of the order of 1.5 mK. But at 0.5 K the *V*(*t*) data do not show TN in many cases (Fig. 4d). The *I-V* curve in the former case also exhibit voltage oscillations with respect to $I_{exc}$ but not in the latter case. When *I-V* curves do not show voltage oscillations the junction is in a stable and higher-$I_c$ state (e.g. Fig. S3 and Fig. 4d). In this case, the system is found to be rather stable even with larger temperature variations. The *V*(*t*) data at 0.5 K in Fig. S5a present the stable state with temperature variations of the order of 50 μK. The *V*(*t*) data exhibit essentially the same constant voltage even with temperature variations of the order of 4 mK (Fig. S5b).

**Possible origin of the switching - ordinary vortex motion**

The ordinary vortex dynamics can cause TN with the motions of the vortices from the Ru/SRO interface to the bulk SRO, or vice versa. In this context the TN may be explained with pinning and depinning of the vortices, but it is difficult to explain the anomalous hysteresis. Furthermore, the effect of ordinary vortices, if playing a main role, should be initiated with the onset of the 3-K superconductivity or at least at 1.8 K where $I_c$ emerges. In contrast, the observed TN is strictly connected with the bulk superconductivity in SRO. In addition, we emphasize that we tried to minimize the effect of trapped vortices by cooling the junctions rather slowly (1 K/hour) in a zero field environment prepared by magnetic shielding.

**Relation between the real-space angle and superconducting phase**

We also confirmed analytically that maximum $I_c$ for given $\theta_{DW}$ is realized at $\alpha(M) = -\alpha(F)$, which is the most stable domain walls configuration. Figure S6 represents the superconducting phase $\varphi$ of $Sr_2RuO_4$ and local critical current $I_{co}\sin\varphi$ as a function real-space angle $\theta$ around Ru inclusion for two different positions of chiral-DW "M" at $\theta_{DW} =$

π/2 and $\theta_{DW} = \pi$, calculated based on the model described in the main text. For $\theta_{DW} = \pi/2$, negative local $I_c$ in certain ranges of $\theta$ reduces the total $I_c$, which is given by the integration around the Ru/SRO circumference. Maximum total $I_c$ is realized for $\theta_{DW} = \pi$, forwhich the negative contribution is absent.

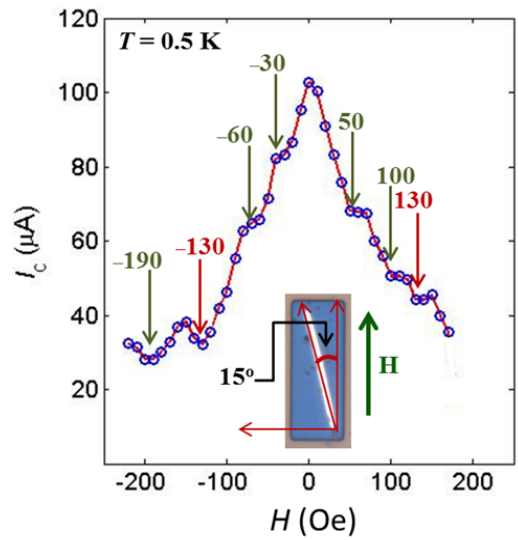

**Figure S1| Critical current versus external applied field.** Fraunhofer pattern with field applied along the *ab*-plane for Junction B at 0.5 K. It shows a clear dips at 130 Oe with additional shoulders at lower field.

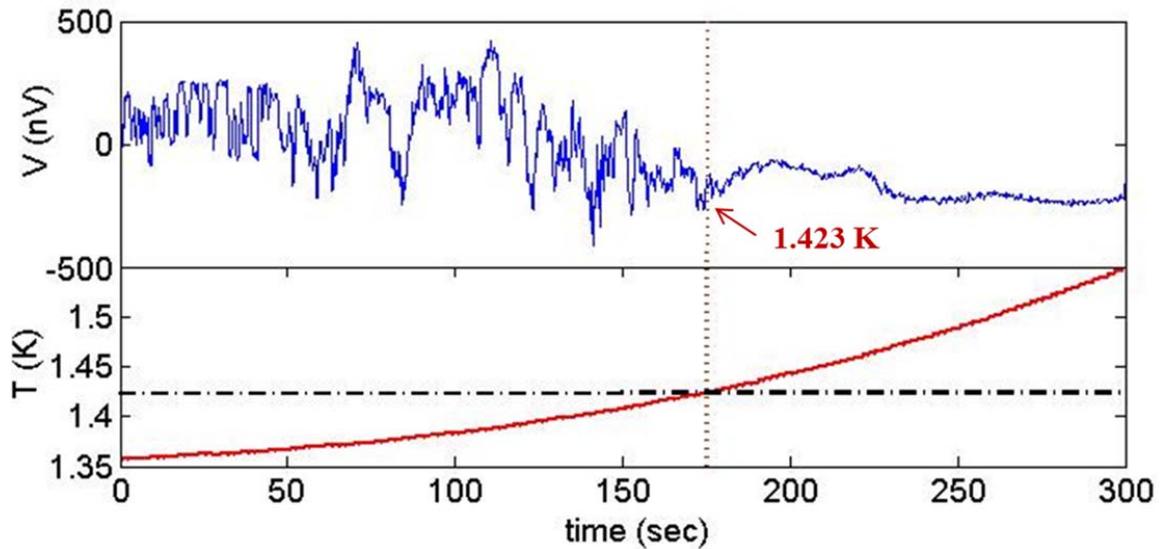

**Figure S2| Temperature dependence of voltage oscillations.** Voltage oscillations versus temperature for Junction A. Upper panel shows the switching of voltage in higher and lower voltage states and lower panel illustrates the increase in temperature as a function of time. The oscillations are totally suppressed just at 1.423K ($T_{c\_bulk}$ of SRO).

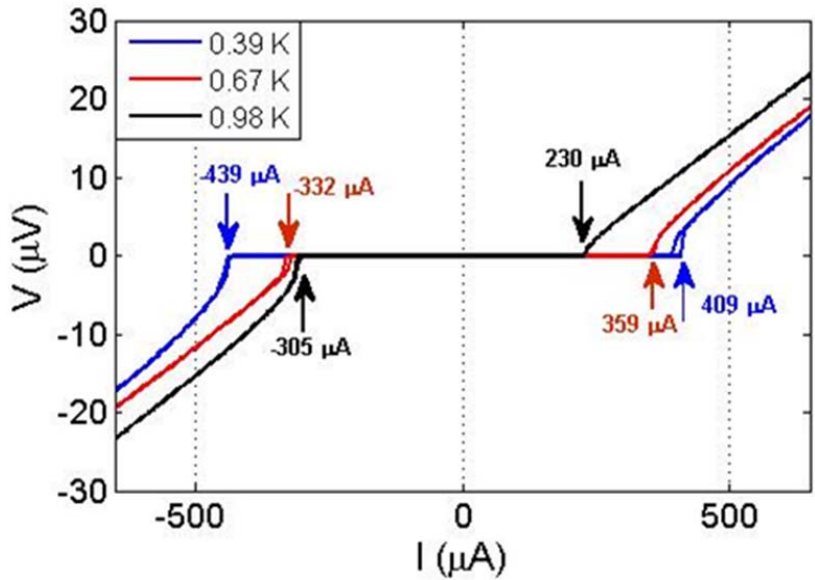

**Figure S3| Asymmetric *I-V* curves.** Asymmetric *I-V* curves with the direction of the current at different temperatures for Junction A. The difference between $I_c+$ and $I_c-$ is obvious with $\Delta I_c = 30$ μA(blue), 27 μA(red), 75 μA(black). During other thermal cycles, we also observed the telegraphic noise in this temperature range.

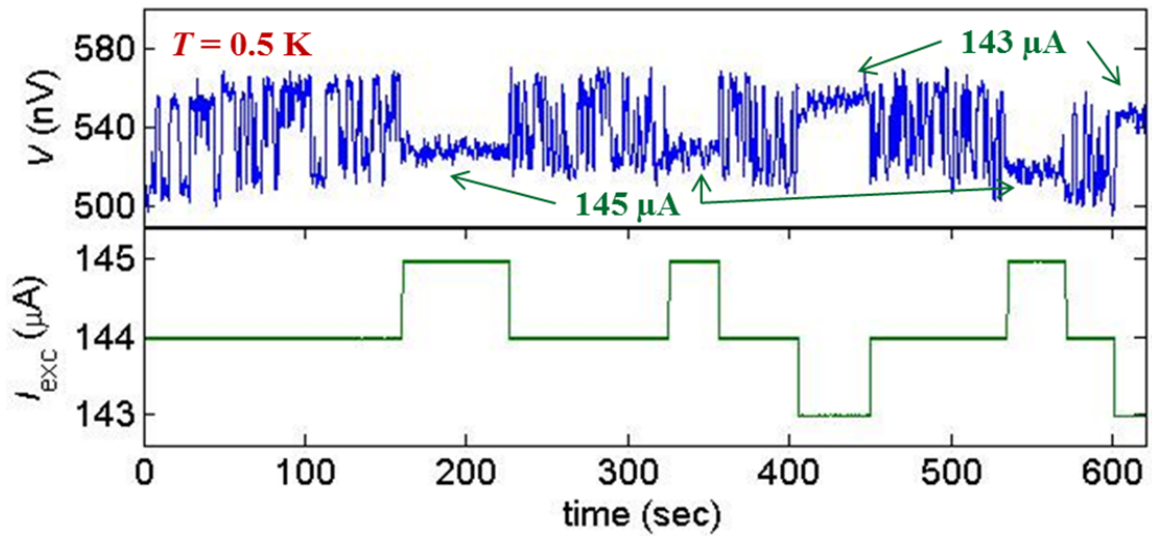

**Figure S4| Switching at $I_{exc} > I_c$ and 0.5 K.** Telegraphic noise for Junction B at different $I_{exc}$ values. Telegraphic noise is present at $I_{exc}$ = 144 µA with the voltage amplitude of ~60 nV. But the voltage turns to be constant just with a change of ±1 µA in $I_{exc}$ (145 µA and 143 µA). Note that the junction voltage at 145 µA is lower than that of at 143 µA. It opposes the Ohmic behavior. This behavior implies a subtle instability to govern the critical current in this junction.

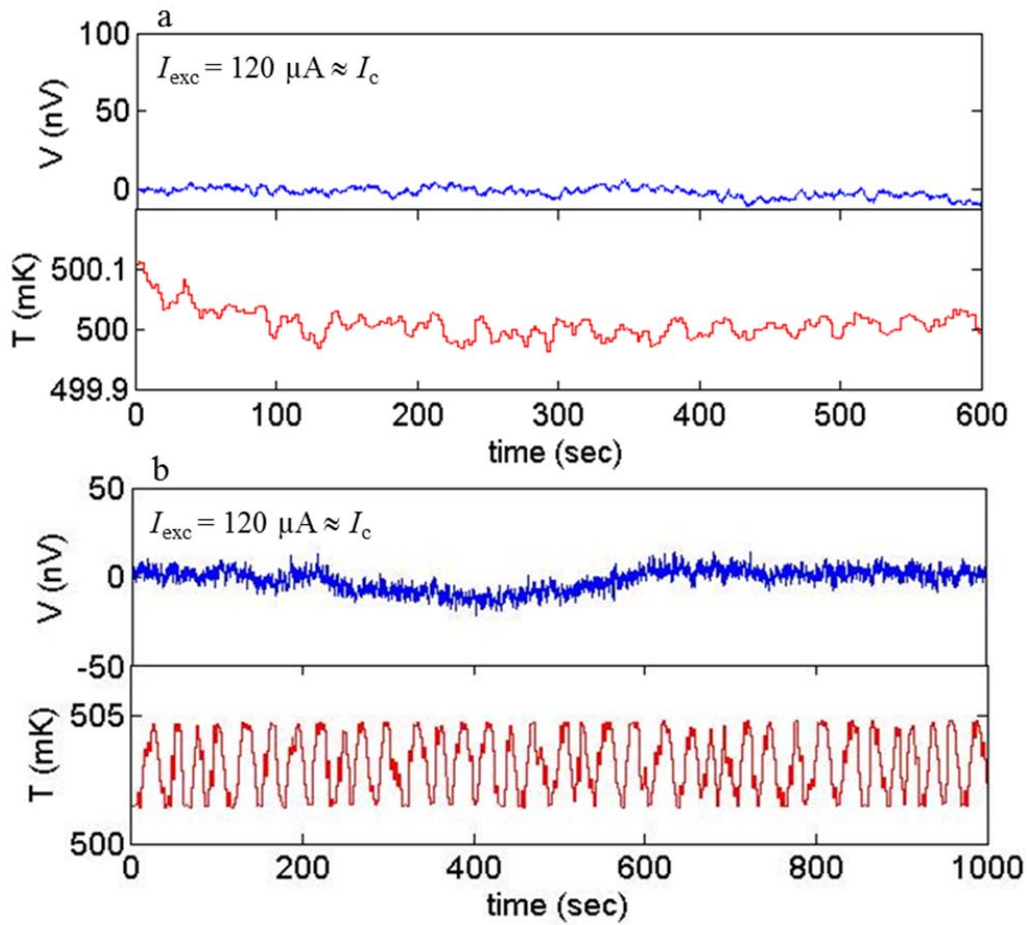

**Figure S5| Voltage switching dependence on temperature variations. a**, Voltage versus time at 0.5 K for Junction A. It illustrates zero voltage without any switching with very stable temperature. **b**, Voltage versus time with much larger temperature variations. It shows again the constant voltage although the temperature variations are of the order of 4 mK.

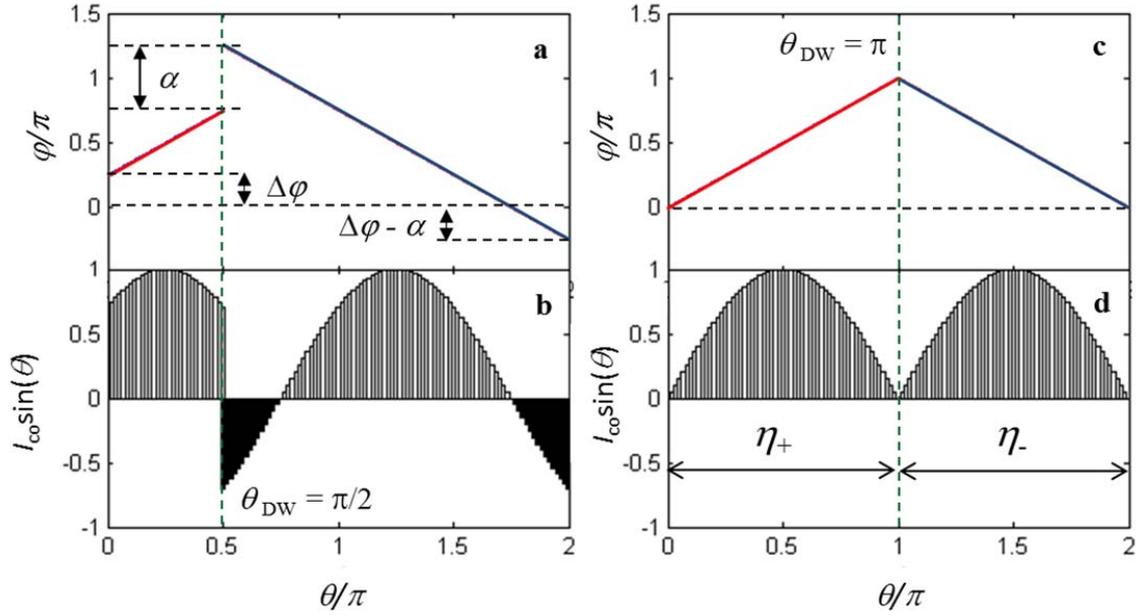

**Figure S6| Superconducting phase and local critical current versus position around Ru inclusion. a,** Superconducting phase $\varphi$ of chiral $p$-wave superconductivity for $\theta_{DW} = \pi/2$ with the phase jump $\alpha$ across the chiral-DW. The red and blue lines represent the phase in the positive and negative chiral domains, respectively and vertical green dotted line represents the value of $\theta_{DW}$. **b,** Local critical current versus angle $\theta$ for $\theta_{DW} = \pi/2$. Positive and negative local $I_c$ is presented by gray area and black area, respectively. **c,** Superconducting phase $\varphi$ as a function of $\theta$ for $\theta_{DW} = \pi$ and $\alpha = 0$. **d,** Local critical current versus angle $\theta$ for $\theta_{DW} = \pi$. Maximum total $I_c$ is realized for this configuration.